\documentclass[prl,amsmath,twocolumn,showpacs]{revtex4-1}
\usepackage{amsmath,graphicx}
\usepackage{epsfig}
\begin{document}
\def\la{{\langle}}
\def\ra{{\rangle}}
\def\a{{\alpha}}
\def\ah{\hat{A}}
\def\h{\hat{H}}
\def\q{{\quad}}

\title{Interference mechanism of seemingly superluminal tunnelling}
%
\author {D. Sokolovski$^{1,2}$ and E. Akhmatskaya$^3$}
\affiliation{$^1$ Department of Chemical Physics,  University of the Basque Country, Leioa, Bizkaia, Spain,\\
$^2$ IKERBASQUE, Basque Foundation for Science, 48011, Bilbao, Spain.\\
$^3$Basque Center for Applied Mathematics (BCAM),\\ Alameda de Mazarredo, 14,  48009 Bilbao, Bizkaia, Spain }
\begin{abstract}
Apparently 'superluminal' transmission, e.g., in quantum tunnelling and its variants, occurs via  a subtle interference mechanism which allows reconstruction of the entire spacial shape of a wave packet from its front tail. It is unlikely that the effect could be described adequately in simpler terms.

\end{abstract}

%
%
\pacs{ 03.65.Ta, 73.40.Gk}
\maketitle
\section{Introduction}
In the early 1930's MacColl \cite{MCOLL} noticed that quantum tunnelling appears to take no time or little time, in the sense that the peak of a
wavepacket,
transmitted across a classically forbidden region, may arrive at a detector earlier than that of the freely propagating one. If the advanced peak is used to predict the time $\tau$ the particle has spent in the barrier region, the result is nearly zero. Dividing the barrier width by $\tau$ yields a velocity exceeding the speed of light $c$, suggesting that the transmission has a 'superluminal' aspect. 
The effect has been predicted and observed for various systems such as potential barriers, semi-transparent mirrors, refraction of light and microwaves in undersized wave guides (for a review see Refs. \cite{REV}-\cite{REV3}). Since below a barrier evanescent  waves decay, rather than propagate, it cannot be explained in terms of superluminal group velocities found, for example,  in propagation in transparent media with inverted atomic populations \cite{CHIAO}
\newline
\section{The search for a physical mechanism}
One outstanding question concerns the physical mechanism of the superluminal effect in tunnelling.
Several authors have made efforts toward answering it.
Nimtz and co-workers suggested that 'superluminality' in propagation of electromagnetic pulses could be explained in terms of virtual photons capable of violating Einstein relativity on a microscopic scale \cite{NIM1}-\cite{NIM4}. 
Winful, \cite{REV3}, \cite{WIN1}-\cite{WIN3}, refused to take the claim of violation of special relativity seriously \cite{WIN3}, since the evanescent wave are described by classical Lorentz-invariant Maxwell equations. Instead he argued that the effect could be explained in terms of the energy (or probability) stored with exponentially decaying density in the classically forbidden region where no actual propagation of the pulse occurs. 
Rather, the energy output at the right end of the barrier adiabatically follows the input at its left end.
Buettiker and Washburn  \cite{BUTT1} dismissed the speculation about superluminal velocities by pointing out, following Refs.  \cite{RESH1} and \cite{JAPHA}, that the transmitted pulse is shaped out of the front end of the incident one, in a manner similar to what is shown in Fig.1.
Winful disagreed \cite{WIN2}, stating 'irreconcilable differences' between his mechanism and the reshaping model. 
He argued that by carving the front part of a two-humped pulse one should get a single-humped transmitted pulse, whereas what goes through in an experiment repeats the original two-humped shape (see the diagram in Fig.1).  This controversy, to our knowledge still unresolved \cite{BUTT2}, is the subject of this paper. 
We show that the essential physics is contained in a simple interference effect,
and offer an explanation which doesn't involve, explicitly or implicitly, 
temporal duration of a tunnelling process.
\begin{figure}
\includegraphics[width=5.5cm, angle=0]{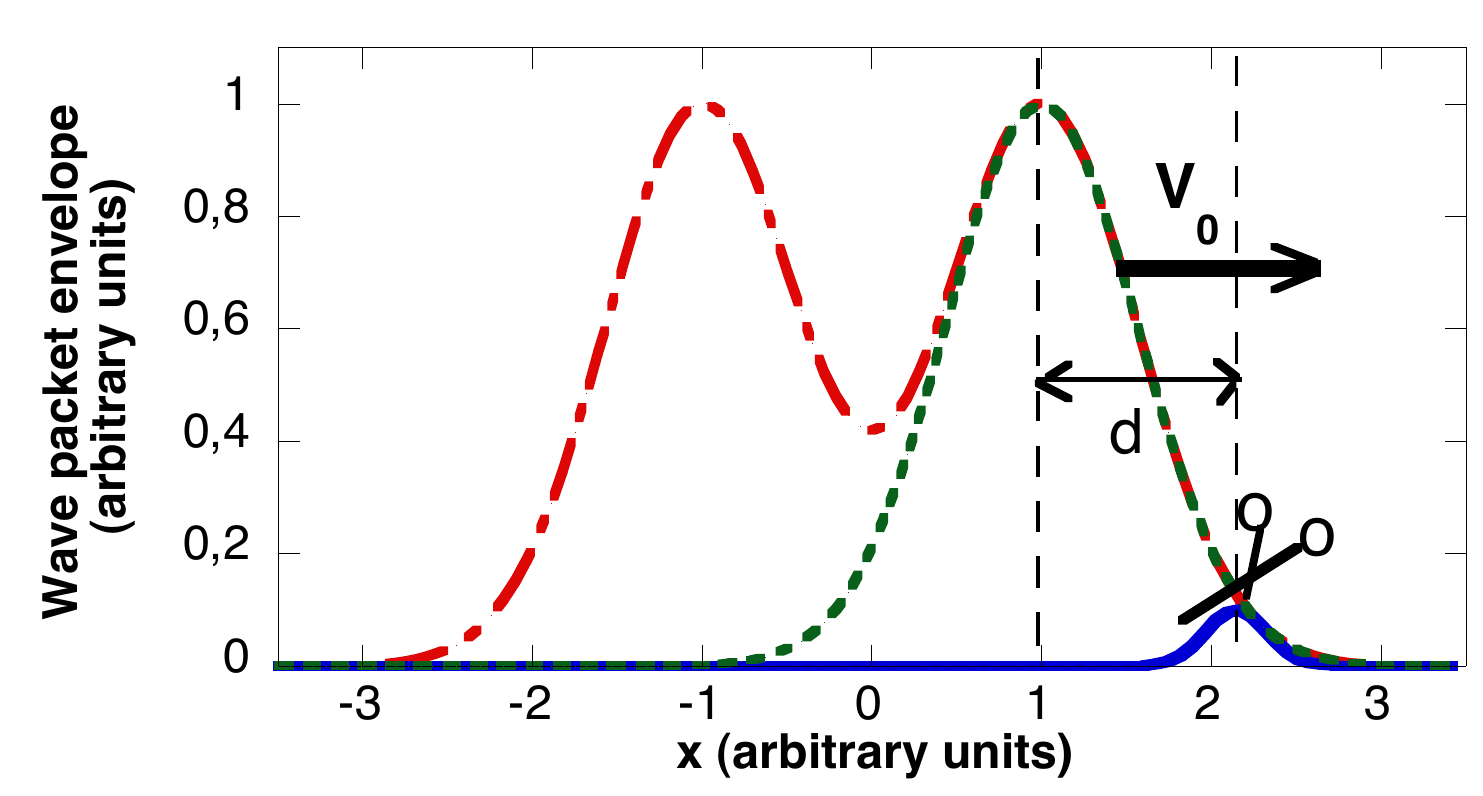}
\label{PROB}
\caption{(colour online) A naive view of reshaping: the transmitted pulse (solid) is carved from 
the front of the incident Gaussian envelope (dashed), whereby its peak is instantly advanced by a distance $d$. The problem is that if the Gaussian is replaced by two-hump shape (dot-dashed), the transmitted pulse should remain single-peaked, and this contradicts the experiment.
}
\end{figure} 
\section{Interference behind the apparently superluminal advancement}
The essence of our analysis is captured by a simple model \cite{DS1} shown in Fig.2. A particle of a unit mass 
carries a large magnetic moment (spin) of $(2K+1)$ components. The particle is described by a wave packet with a  mean momentum $p_0$, and the spin is first prepared (pre-selected) in a state of our choice, $|a\ra =\sum_{m=-K}^K a_m|m\ra$,
and later post-selected in another state $|b\ra =\sum_{m=-K}^K b_m|m\ra$. In between, the particle passes through a region of a width $d$ which contains a small magnetic field. There the $m$-th component of the spins encounters a small rectangular potential $\omega_L m$, where $\omega_L$ is the Larmor frequency. Depending on $m$, the potential slows the particle down, or speeds it up. One can choose $|a\ra$ in such a way that none of the components are sped up.
If the particle is fast, one can also  neglect both the reflection from the edges of rectangular steps and the spreading of the original pulse. 
Then, once the spin is post-selected, (e.g., by passing through a polariser), the transmitted Gaussian  pulse is sum of Gaussians, all (except for one with $m=0$) delayed relative to the free propagation, (we use $\hbar=1$),  
\begin{eqnarray}\label{1}
\Psi(x,t)=\exp(ip_0x-p_0^2t/2)G^T(x,t),\q\q\q
\end{eqnarray}
\begin{eqnarray}\label{2}
G^T(x,t) = N^{-1/2} \sum_{m=0}^K\eta_m G_0(x-p_0t+m\Delta x),
\end{eqnarray}
\begin{eqnarray}\label{3}
  \eta_m\equiv \exp(-im\omega_Ld/p_0)a_mb^*_m,
\end{eqnarray}
\begin{eqnarray}\label{4}
G_0(x)=(2/\pi \sigma^2)^{1/4} \exp(-x^2/\sigma^2),
\end{eqnarray}
where 
 $\Delta x=\omega_Ld/p_0^2$ and $\sigma$ is the pulse's width. We also
assumed that 
$a_m$ and $b_m$ may take arbitrary values, and introduced the normalisation factor $N\equiv \la a|a\ra\la b|b\ra$.
\begin{figure}
\includegraphics[width=5.5 cm, angle=0]{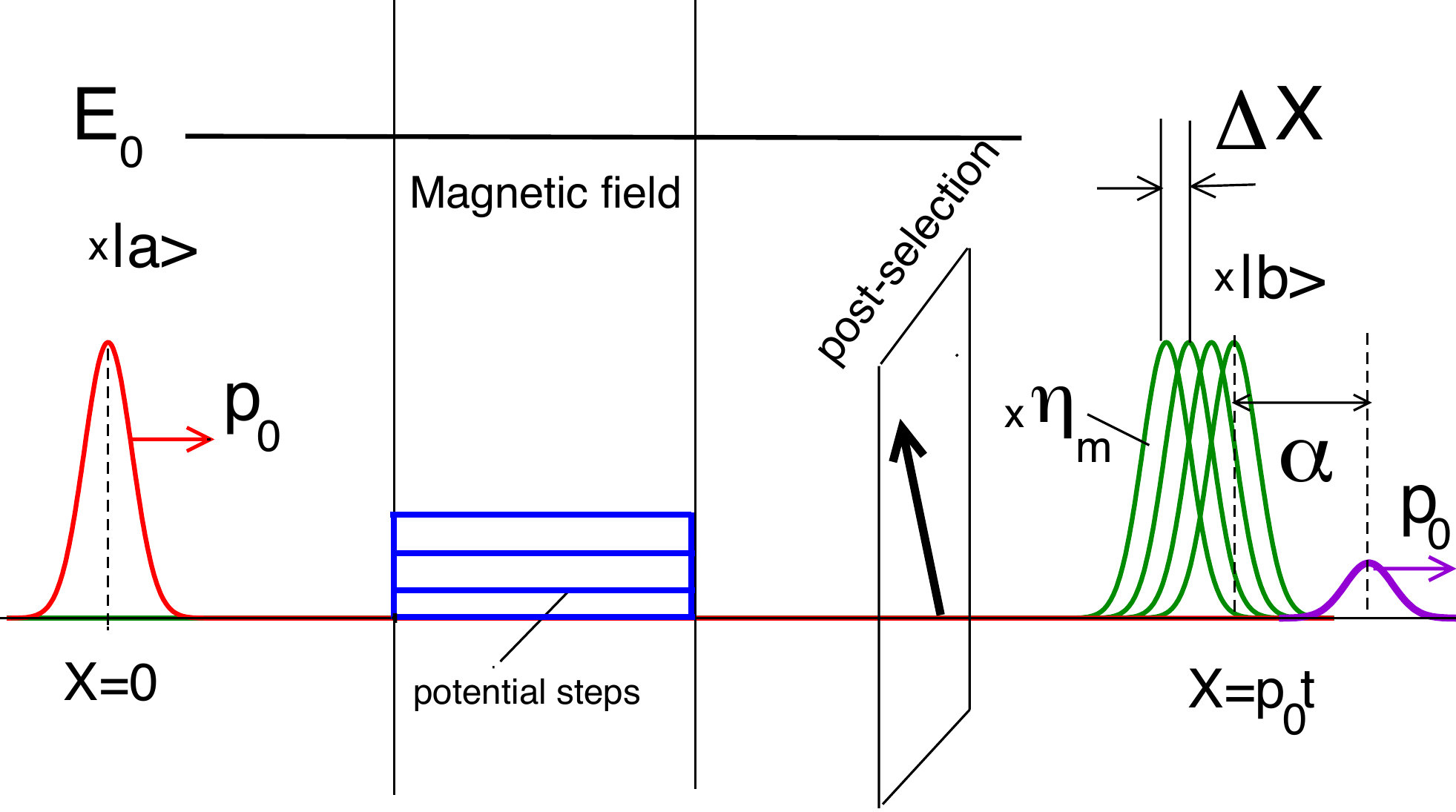}
\label{PROB}
\caption{(colour online) Schematic diagram illustrating our model. Large spin of a fast particle is pre-selected in a state $|a\ra$ prior to entering a magnetic field in which all spin components are delayed. The
spin state is purified upon passing the polariser, after which the coordinate part of the wavefunction is given by a superposition of delayed pulses  weighed by complex quantities $\eta_m$.}
\end{figure}  
With the technical details explained, we can focus on our main interest, 
which is to design the shape of the transmitted pulse by choosing appropriate $a_m$ and $b_m$.
As the title suggests, we may want to put it a distance $\alpha = d$ ahead of the freely propagating wave packet, making it look like the field was crossed infinitely fast. Or even by $n$ times the distance $d$, 
where the same logic would lead us to a negative duration spent in the field. 
(This alone might put one off the idea to describe the transmission in terms of 'transmission times'.)
With $K$ sufficiently large, our goal can be achieved by choosing $\eta_m$ to satisfy
\begin{eqnarray}\label{5}
\sum_{m=0}^K (m\Delta x)^j\eta_m/\sum_{m=0}^K \eta_m = (-nd)^j, \q 0\le j \le K\q
\end{eqnarray}
Then, expanding the Gaussians in Eq.(\ref{2}) around $x=p_0t$ in a Taylor series, and re-summing the series, we have the desired result $G^T(x,t)\approx CG(x-p_ot-nd)$, where $C=\sum_{m=0}^K \eta_m/N^{1/2}$. 
Analytical expressions for $\eta_m$ satisfying (\ref{5}) are known from Ref.\cite{DS1}, where it was also shown that the factor $C$ rapidly decreases for larger displacements $nd$, as successful post-selection in $|b\ra$ becomes less probable. 
Since our discussion is one of principles, this is sufficient for a comparison with the conclusions of Refs. \cite{REV3}, \cite{WIN1}-\cite{WIN3} and \cite{BUTT1}.
\newline
Our advancement mechanism is of interference nature, similar to the one known in the 'weak' quantum measurements \cite{Ah}, \cite{DS3}. 
The setup in Fig.2  splits the incident envelope into $K$ delayed copies, which interfere destructively everywhere except in the far right, where their front tails combine to produce a reduced copy of the original pulse. 
The transmitted pulse is, indeed, 'front loaded' as suggested in Ref.\cite{BUTT1}, yet the mechanism is not just the primitive reshaping shown in Fig.1. The device in Fig.2
reconstructs the global structure of an analytical function (the pulse's envelope) from 
the information contained locally in its front tail. It is also a linear device, and a two-hump shape shown in Fig.1 is reshaped into a two-hump transmitted pulse, as is shown in Fig.3a. The fact that the task becomes more difficult further away from the main structure of the original pulse, explains the rapid decrease in the post-selection success rate for large advancements. The energy, or probability  storage explanation of Refs. \cite{REV3}, \cite{WIN1}-\cite{WIN3} clearly does not apply, since neither is stored anywhere in the setup. 
\section{The speed of information transfer}
It is easy to demonstrate that the technique can be used for fast, but not 'superluminal' communication, e.g., by encoding $0$ and $1$  in single- and double-hump pulses respectively. By using a setup similar to the one shown in Fig.2 one would we able to distinguish between $0$ and $1$ sooner, than if the carrier pulse were sent by free propagation. There is, however, nothing 'superluminal' about this early detection, as the device only interprets the small amount of information arriving to the detector via causal 'subliminal' route. Suppose, for example, that the sender turns a two-hump pulse into a single-hump one by cutting it down the middle and discarding the rear part,
yet leaving the front tail untouched. Then the advanced part of the transmitted pulse will reproduce the now non-existent two-hump structure, as shown in Fig.3.  The recipient will not realise his/her mistake until he/she receives the 'subluminal' part of the signal, produced where the destructive interference would make the uncut signal vanish. This example also contradicts the view \cite{WIN2} that the output 'adiabatically follows the input', since there is no second hump in the input.
\newline
The analysis can also be carried out in the momentum space.
 The transmitted amplitude can equivalently  be written as (we introduce $X\equiv x-p_ot$ to shorten notations)
\begin{eqnarray}\label{6}
G^T(X) = N^{-1/2}\int T(p)A(p)\exp(ipX)dp,
\end{eqnarray} 
where $G(x)=\int A(p)\exp(ipx)dp$ and the transmission amplitude  $T(p)$ is the  Fourier transform of  the $\eta_m$,
\begin{equation}\label{7}
T(p) \equiv \sum_{m=0}^K \eta_m \exp(imp\Delta x).
\end{equation} 
With $\eta_m$ determined by Eq.(\ref{5}), the transmission amplitude develops a well defined  'window' or 'band', inside which $T(p)$, although built from exponentials with only non-negative frequencies $m\Delta x\ge 0$, is proportional to $ \exp(-ipnd)$.
 This is an example of 'super-oscillations', a term coined by Berry \cite{BERRY1} 
in order to describe local oscillations of a function with a frequency outside its Fourier spectrum.
Thus, any pulse with a momentum distribution narrow enough to fit into this super-oscillatory window will be accurately advanced. A spacial shift by, say, $y$ of a pulse  as a whole does not broaden its momentum distribution $A(p)$, but only multiplies it by $\exp(-ipy)$. Thus, any initial pulse of the form $ \sum_j G(x-y_j)$, single-, double- or multi-humped, will be advanced as a whole, to lie ahead of its freely propagating counterpart. 
\begin{figure}
\includegraphics[width=8.5cm, angle=0]{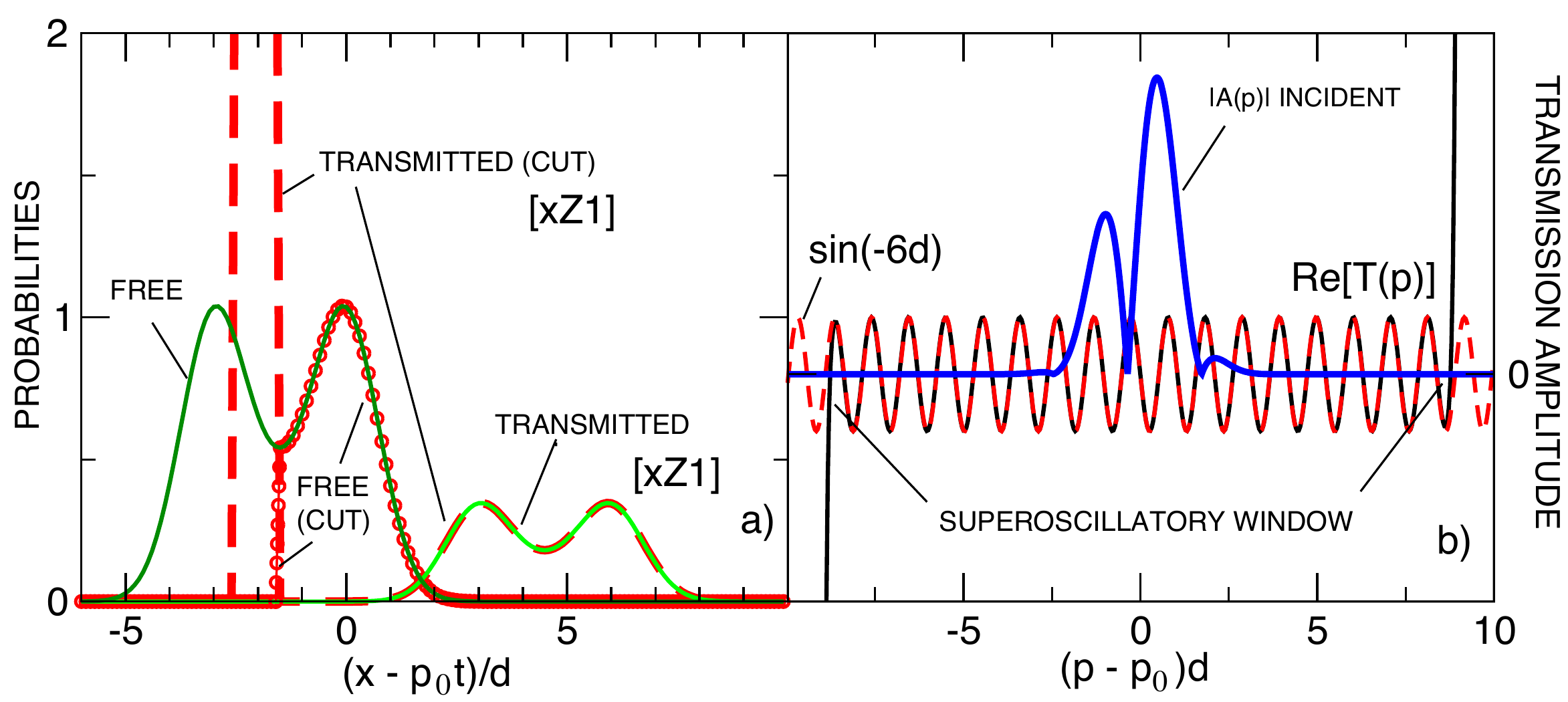}
\label{PROB}
\caption{(colour online) a) A two-hump pulse transmitted by a setup shown in Fig.2 (solid) (multiplied by a large factor $z1=3.3*10^{461}$ for better viewing \cite{BAIL}), and the same initial pulse evolved by free motion. Also shown is the free evolution of the same pulse cut down the middle with the rear part amputated (circles). In this case, the transmitted pulse has the same advanced part followed by a large delayed signal (dashed).
b) Superoscillatory window in the transmission amplitude and the momentum distribution of the initial two-hump pulse. The parameters are $n=6$, $K=150$, and $\sigma/d=1.5$.  }
\end{figure}
\section{The interference mechanism of apparently superluminal tunnelling}
Next we show that essentially the same mechanism, albeit without 
the flexibility in choosing a desired advancement, is employed in  quantum tunnelling and its variants, such as transmission of electromagnetic waves in undersized wave guides. 
Although the physical setup of a tunnelling experiment is clearly different from the one shown in Fig .2,
we readily recover the analogues of Eqs. (\ref{2}), (\ref{6}) and (\ref{7}).
The transmitted wave packet has a form similar to Eq.(\ref{6}) (we put the particle's mass to unity),
\begin{eqnarray}\label{8}
\Psi^T(x,t) =\int T(p) A(p-p_0)\exp(ipx-ip^2t/2) dp,
\end{eqnarray} 
where $T(p)$ is the barrier's transmission amplitude, and $A(p-p_0)$, peaked at the mean momentum $p_0$, is the momentum distribution. If the potential does not support bound states, 
the Fourier spectrum of $T(p)$ cannot contain negative frequencies, and we have an analogue of Eq.(\ref{7})
 \begin{eqnarray}\label{9}
T(p)=\int_0^{\infty}\exp(ipx)\xi(x)dx,
\end{eqnarray} 
where $\xi(x)\equiv(2\pi)^{-1}\int_{-\infty}^{\infty}T(p)\exp(-ipx)dp$ vanishes for $x<0$, since $T(p)$ has no poles in the upper half of the $p$-plane. Finally, reverting to the coordinate space by re-writing (\ref{8}) as a convolution yields the analog of Eq.(\ref{2}),
\begin{eqnarray}\label{10}
\Psi^T(x,t) =\int dx' \xi(x') \Psi_0(x-x',t),
\end{eqnarray}
where $\Psi_0(x,t)=\int A(p-p_0)\exp(ipx-ip^2t/2) dp$ is the freely propagating state, i.e., what the initial wave packet would have evolved into by the time $t$,  had there been no barrier. As 
in Eq.(\ref{2}),
 the transmitted pulse results from the interference between the freely propagating pulse and its delayed copies. 
\newline
As in our first example, we have evidence of superoscillatory behaviour.
Consider tunnelling across
 a broad rectangular barrier of a height $V$ and a width $d$, $p_0d>>1$. Expanding the barrier action $d(2V-p^2)^{1/2}$, for $p$ close to $p_0$ we can approximate the transmission amplitude by
\begin{eqnarray}\label{11}
T^{approx}(p)=
T(p_0) \exp[-i\a (p-p_0)+\beta (p-p_0)^2],\q 
\end{eqnarray}
where $\a =  d+ip_0d/(2V-p_0^2)^{1/2}$ and $\beta=Vd/(2V-p_0^2)^{3/2}$.
The factor $\exp(-ipd)$, which suggests advancement of the transmitted pulse by the barrier length $d$, is clearly 'super-oscillatory', since there are no negative frequencies in the Fourier transform (\ref{9}).
\newline
To observe the advancement we require an incident pulse so broad that the approximation (\ref{11}) would hold for all its momenta.
A Gaussian wave packet $\Psi^G$, initially centred at some $x_0$, evolves by free motion into
\begin{eqnarray}\label{12}
\Psi^G_0(x,t|\sigma,x_0)=\exp(ip_0x-ip_0^2t/2)
\times\q\q\q \q\q\q  \\ \nonumber
[2\sigma^2/\pi\sigma_t^4]^{1/4}\exp[-(x-p_0t-x_0)^2/\sigma_t^2],
\end{eqnarray}
where $\sigma_t=\sqrt{\sigma^2+2it}$ accounts for the broadening of the pulse.
We may want to send a two-hump pulse composed of two shifted Gaussians, 
$\Psi_0(x,t=0)=\Psi^G_0(x,0|\sigma,0)+\Psi^G_0(x,0|\sigma,-\sigma/2)$. From Eqs.
(\ref{8}) and (\ref{11}), we expect the transmitted pulse to undergo a reduction, a shift into the complex coordinate plane by $Re\a+iIm\a$, and an additional broadening proportional to $\beta$, 
 \begin{eqnarray}\label{14}
\Psi^T(x,t) = T(p_0)[\Psi^G_0(x-\a,t|\sqrt{\sigma^2-4\beta},0)+ \q\q\q\\ \nonumber
\Psi^G_0(x-\a,t|\sqrt{\sigma^2-4\beta},-\sigma/2)].
\end{eqnarray} 
In the case shown in Fig. 4, the exact result coincides with Eq.(\ref{14}) to a graphical accuracy. The transmitted pulse is two-humped, with the humps advanced  by $d$ relative to free propagation. As in the first example, the interference mechanism, linear in the input, detects two Gaussian front tails and faithfully reconstructs each of the constituent Gaussians.
Provided the width of the pulse is adjusted, one can 
observe the same effect for ever broader barriers \cite{DS3}. In this way, the interference mechanism provides an explanation for the Hartman effect \cite{REV3}: the interference puts the transmitted pulse where it looks as if it has spent zero time in an arbitrary broad classically forbidden region. 
As in the first example, this analysis does not employ the concept of  a duration spent inside the barrier or, indeed, any other tunnelling time mentioned \cite{BUTT1}. 
\begin{figure}
\includegraphics[width=8.5cm, angle=0]{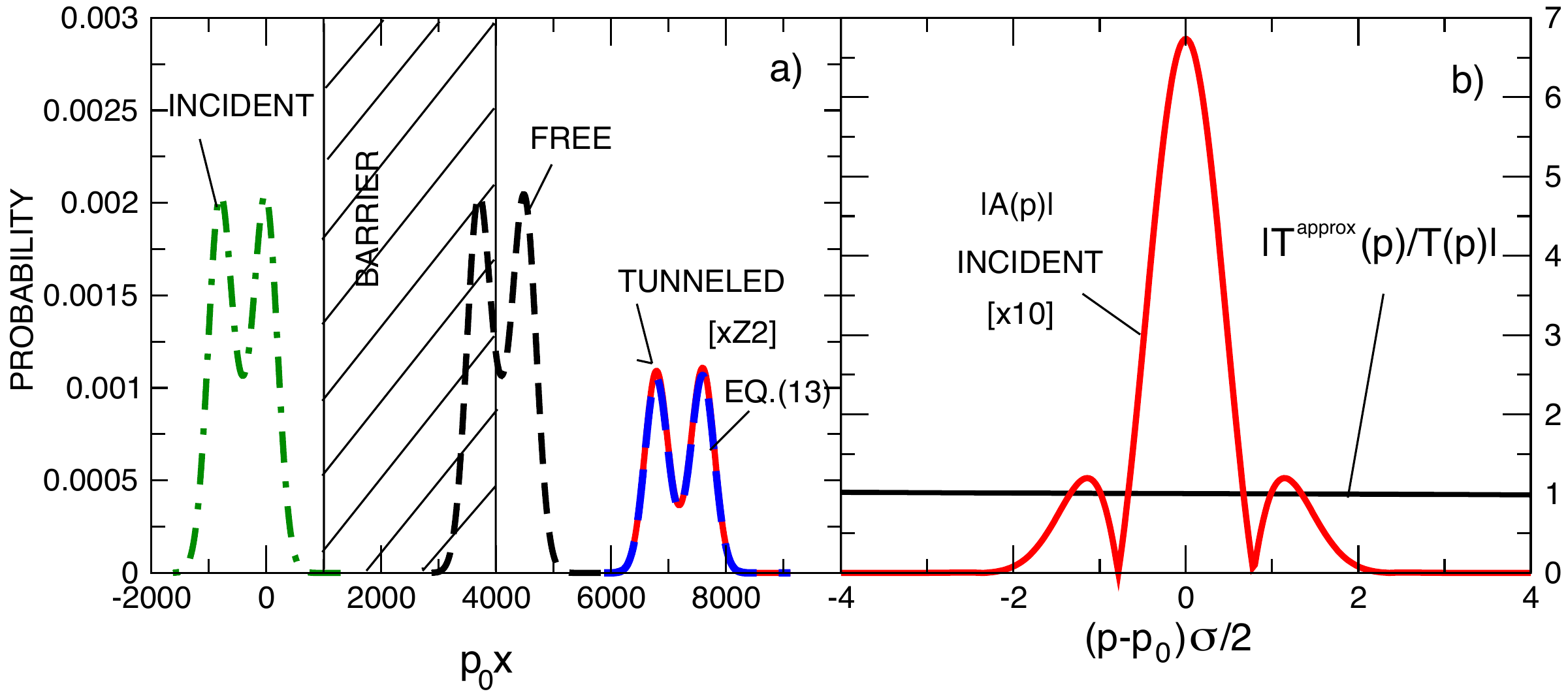}
\label{PROB}
\caption{(colour online) a) A two-hump pulse tunnelled across a broad rectangular barrier (solid) (multiplied by a large factor $Z2=4*10^{14}/|T(p_0)|^2$ for better viewing), the same initial pulse evolved by free motion (dashed), and the incident pulse at $t=0$ (dot-dashed).
b) Transmission amplitude and the momentum distribution of the initial two-hump pulse (multiplied by a factor of 10). The parameters are $p_0d=3000$, $V/p_0^2=2$, $t=1.5d/p_0$, $\sigma/d=0.135$. }
\end{figure}
\section{Conclusions}
In summary, the 'superluminal' transmission occurs via a subtle interference mechanism. One valid analogy is that of a beam splitter which splits the initial pulse into many components, and where nothing moves faster than in free propagation. 
The amplitudes and phases attached to each components are such that, when recombined, they cancel each other everywhere, except in the forward region.
This gives the advanced transmitted wave packet a 'superluminal' aspect. The necessary and sufficient condition for the existence of the effect is that all incident momenta must probe local 'super-oscillatory' behaviour of the transmission amplitude. This can be true for a broad analytical pulse, in which case the entire shape is reconstructed from the information contained in its front tail \cite{FOOT}. 
The presence of a non-analytical feature, such as a sharp cut-off, 
inevitably broadens the momentum distribution, and destroys the effect.
\newline
Thus, we agree with Ref. \cite{BUTT1} that the emerging pulse is 'front-loaded', 
with an addition that, unlike the primitive reshaping shown in Fig.1, the interference mechanism 
must turn a multi-hump combination of initial Gaussians into a multi-hump transmitted pulse since transmission is linear in the input.  
We find no evidence to support the energy storing mechanism where 'the output adiabatically follows the input' \cite{WIN2}. One reason is that 'superluminality' may be achieved in a system where no energy storage occurs, e.g, in the one shown in Fig.2. Another is that the advanced field may persist even if with the rear part of the incident pulse modified or amputated, i.e., in the case where there is no 'input' to follow.  
\newline
We stress the convenience of analysing the spacial shape of the transmitted pulse, rather than the time variation of the signal at a fixed location. Once this shape is known, one can evaluate the physically meaningful times, such as the arrival of the pulse's peak at a remote detector.  This, in turn, allows one to avoid the notion of suspiciously short 'tunnelling times' and disproportionately high 'tunnelling velocities' which have plagued the subject from its inception in the early 1930's.
\section{Acknowledgements}
One of us (DS) acknowledges support of the Basque Government Grant No. IT472 and MICINN (Ministerio de Ciencia e Innovacion) Grant No. FIS2009- 12773-C02-01.


\begin{thebibliography}{999}
\bibitem{FOOT} Note that exact reconstruction of the pulse is possible for the setup 
shown in Fig. 2. In the case of tunnelling, an additional deformation caused by the complex shift
by $iIm \a$ remains even for a pulse so broad that the quadratic term in the exponent of (\ref{11})
can be neglected.
\end{thebibliography}
\end{document}